\documentclass[runningheads]{llncs}
\usepackage{graphicx}
\usepackage[font=small]{subfig}
\usepackage[compatible]{algpseudocode}
\usepackage{algorithm}
\usepackage{epstopdf}
\usepackage{multirow}
\begin{document}
\captionsetup[subfigure]{labelfont=scriptsize,textfont=scriptsize}

\title{Constructive Heuristics for Min-Power Bounded-Hops Symmetric Connectivity Problem\thanks{ The research is supported by the Russian Science Foundation (project 18-71-00084).}}
\titlerunning{Constructive Heuristics for MPBHSCP}
\author{Roman Plotnikov\inst{1}\orcidID{0000-0003-2038-5609} \and
Adil Erzin\inst{1,2}\orcidID{0000-0002-2183-523X}}
\authorrunning{R. Plotnikov et al.}

\institute{Sobolev Institute of Mathematics, Novosibirsk, Russia \and
Novosibirsk State University, Novosibirsk, Russia}
\maketitle              % typeset the header of the contribution
\begin{abstract}
We consider a Min-Power Bounded-Hops Symmetric Connectivity problem that consists in the construction of communication spanning tree on a given graph, where the total energy consumption spent for the data transmission is minimized and the maximum number of hops between two nodes is bounded by some predefined constant. We focus on the planar Euclidian case of this problem where the nodes are placed at the random uniformly spread points on a square and the power cost necessary for the communication between two network elements is proportional to the squared distance between them. Since this is an NP-hard problem, we propose different polynomial heuristic algorithms for the approximation solution to this problem. We perform a posteriori comparative analysis of the proposed algorithms and present the obtained results in this paper.
\end{abstract}
\section{Introduction}
Due to the prevalence of wireless sensor networks (WSNs) in human life, the different optimization problems aimed to increase their efficiency remain actual. Since usually WSN consists of elements with the non-renewable power supply with restricted capacity, one of the most important issues related to the design of WSN is prolongation its lifetime by minimizing energy consumption of its elements per time unit. A significant part of sensor energy is spent on the communication with other network elements. Therefore, the modern sensors often have an ability to adjust their transmission ranges changing the transmitter power. Herewith, usually, the energy consumption of a network's element is assumed to be proportional to $d^s$, where $s\ge 2$ and $d$ is the transmission range \cite{R96}.

The problem of search of the optimal power assignment in WSN is well-studied. The most general Range Assignment Problem, where the goal is to find a strongly connected subgraph in a given oriented graph, has been considered in \cite{CPS99,KKKP00}. Its subproblem, Minimum Power Symmetric Connectivity Problem (MPSCP), was first studied in \cite{CMZ02}. The authors proved that Minimum Spanning Tree (MST) is 2-approximation solution to this problem. Also, they proposed a polynomial-time approximation scheme with a performance ratio of $1 + \ln{2} + \varepsilon \approx 1.69$ and 15/8-approximation polynomial algorithm. In \cite{CNSCL03} a greedy heuristic, later called  Incremental Power: Prim (IPP), was proposed. IPP is similar to the Prim's algorithm of the finding of MST. A Kruscal-like heuristic, later called Incremental Power: Kruscal, was studied in \cite{CN02}. Both of these so-called incremental power heuristics have been proposed for the Minimum Power Asymmetric Broadcast Problem, but they are suitable for MPSCP too. It is proved in \cite{PS06} that they both have an approximation ratio 2, and it was shown in the same paper that in practice they yield significantly more accurate solution than MST. Also, in a series of papers different heuristic algorithms have been proposed for MPSCP and the experimental studies have been done: local search procedures \cite{PS06,ACMPTZ06,EPS13}, methods based on iterative local search \cite{WM09}, hybrid genetic algorithm that uses a variable neighborhood descent as mutation \cite{EP15}, variable neighborhood search \cite{EMP16_COR}, and variable neighborhood decomposition search \cite{PEM18_VNDS}.

Another important property of WSN's efficiency is message transmission delay, i.e., the minimum time necessary for transmitting a message from one sensor to another via the intermediate transit nodes. As a rule, the data transmission delay is proportional to the maximum number of hops between two nodes of a network. The general case, when the network is represented as directed arc-weighted graph, and the goal is to find a strongly connected subgraph with minimum total power consumptions and bounded path length, is called Min-Power Bounded-Hops Strong Connectivity Problem. In \cite{KKKP00} a special Euclidian case of this problem, when equidistant on the line, was considered. In \cite{Clementi00_1} the approximation algorithms with guaranteed estimates have been proposed for the Euclidean case of this problem. The bi-criteria approximation algorithm for the general case (not necessarily Euclidian) with guaranteed upper bounds has been proposed in \cite{Calinescu06}. The authors of \cite{Carmi15} propose improved constant factor approximation for the planar Euclidian case of the problem.

In this paper, we consider the symmetric case of Min-Power Bounded-Hops Strong Connectivity Problem, when the network is represented as undirected edge-weighted graph.  Such a problem is known as Min-Power Bounded-Hops Symmetric Connectivity Problem (MPBHSCP) \cite{Calinescu06}. We also assume that sensors are positioned on Euclidian two-dimensional space. Energy consumption for the data transmission is assumed to be proportional to the area of a circle with center in sensor position and radius equal to its transmission range $d$, and, therefore, $s$ is considered to be equal $2$. This problem is still NP-hard in two-dimensional Euclidian case \cite{Clementi00}, and, therefore, the approximation heuristic algorithms that allow obtaining the near-optimal solution in a short time, are required for it.

Although MPBHSCP is known to be NP-hard, to the best of our knowledge, none research has been done to find the most efficient in practice approximation algorithms. This paper is aimed to fill this gap. We propose six different constructive heuristics for the approximation solution of MPBHSCP. We employ the ideas of the most natural and widely spread heuristics for the Bounded-Diameter Minimum Spanning Tree (BDMST). We conducted an extensive numerical experiment where these algorithms have been compared. We present the results of the experiment in this paper.

The rest of the paper is organized as follows. In Section \ref{sPF} the problem is formulated, in Section \ref{sH} descriptions of the proposed algorithms are given, Section \ref{sS} contains results and analysis of an experimental study, and Section \ref{sC} concludes the paper.

\section{Problem formulation}\label{sPF}

Mathematically, MPBHSCP can be formulated as follows. Given a connected edge-weighted undirected graph $G = (V,E)$ and an integer value $D \geq 1$, find such spanning
tree $T^*$ of $G$, which is the solution to the following problem:

$$
    W(T)=\sum\limits_{i\in V}\max\limits_{j\in V_i(T)}
    c_{ij}\rightarrow\min\limits_T,
$$
$$
    dist_T(u, v) \leq D \; \forall u,v \in V,
$$
where $V_i(T)$ is the set of vertices adjacent to the vertex $i$ in the
tree $T$, $c_{ij} \geq 0$ is the weight of the edge $(i,
j) \in E$ and $dist_T(u, v)$ is the number of edges in a path between the vertices $u \in V$ and $v \in V$ in $T$.

Obviously, in general case, MPBHSCP may even not have any feasible solution. In this paper, we consider a planar Euclidian case, where an edge weight equals the squared distance between the corresponding points and $G$ is a complete graph. Also, we assume that the sensors are randomly uniformly distributed on a square with fixed side. Therefore, for example, the density of a network grows with increase of the number of its elements.

\section{Heuristic algorithms}\label{sH}

We propose a set of heuristic algorithms that construct an approximate solution to the MPBHSCP. Many of them use ideas that previously have been applied to the solution of Bounded-Hops Minimum Spanning Tree (BDMST). As well as it is done in many efficient heuristic algorithms for BDMST, we will use a \emph{center-based} approach, where, at first, the center (one vertex if $D$ is even or two vertices if it is odd) is chosen, and after that, the tree is constructed taking care of the depth of each vertex in relation to the center. The main difference between the algorithms applied for BDMST and our methods is the calculation of the objective function increment after the small modifications of a partial solution. An objective function of MST is additive, that is, adding (or removing) an edge will increase (or decrease) the objective value exactly by the weight of an edge, which is not held for the objective function of MPSCP: if one wants to calculate the change of an objective function value for MPSCP after adding or removing an edge, then he has to take into account the weights of all adjacent edges of a tree.

Let us define the notations that will be used further. For the convenience purposes, we will construct a directed tree, rooted in a center. If the center contains two vertices then one of them will be referred to as a root, and the second one --- as its child. Let's call the minimum number of edges between a vertex and center in a tree as the \emph{depth} of a vertex. Let $V_T \subset V$ stand for a set of vertices in a tree $T$, $E_T$ stand for a set of edges of $T$. Let $Parent_T(v) \in V_T$ be a parent of a vertex $v \in V_T$. If $v \notin V_T$ then let $Parent_T(v) = \emptyset$. Let $depth_T(v)$ be the depth of a vertex $v$ in a tree $T$ in relation to the center, that is the minimum number of hops (edges) between $v$ and center. If $v \notin V_T$ then let $depth_T(v)$ be equal to $-1$. Let $Power(v, u) = Power(u, v)$ be the power cost necessary for the direct communication between the vertices $u$ and $v$. As it was mentioned before, in this paper, we assume that $Power(u,v) = (pos_u - pos_v)^2$, where $pos_u$ and $pos_v$ are positions in Euclidian two-dimensional space of, correspondingly, the vertices $u$ and $v$. Of course, these values may be calculated once since the positions are fixed. $Power_T(u)$ will stand for the power consumptions of a vertex $v$ in a tree $T$. $N_T(v) \subset V_T \{v\}$ will stand for a set of neighbors of $v \in V_T$ in $T$. That is, $Power_T(v) = \sum_{u \in N_T(v)}{Power(u, v)}$, and the total power consumption of a tree $T$, which is the objective function value, is $W(T) = \sum{v \in V_T}{Power_T(v)}$.

\subsection{Prim-Like Heuristics}
Many of known greedy approaches for BDMST use the Prim's strategy \cite{Prim1957} for tree building. Starting from a tree with the only vertex, these algorithms repeatedly add a new edge that connects a non-tree vertex with a vertex in a tree and does not violate the restriction on the diameter. Herewith, criteria of choosing the new non-tree vertex may vary while the in-tree vertex is always chosen greedily. A way of choosing the center vertices, which is rather essential, may vary too. The general scheme of the Prim-Like Heuristic (PLH) is presented in Algorithm \ref{alg:plh}. Below we will consider three different heuristics that are based on the Prim's strategy: Min-Power Center-Based Tree Construction, Min-Power Randomized Tree Construction, and Min-Power Center-Based Least Sum-of-Costs. The difference between these algorithms lies in the different implementations of the methods $ChooseFirstCenters$, $ChooseSecondCenter$, and $ChooseEachVertex$.
\begin{algorithm}
\begin{algorithmic}
\STATE $C[.] \leftarrow ChooseFirstCenters()$;
\STATE $W^* \leftarrow \infty$;
\FORALL {$v_0 \in C[.]$}
\STATE $V_0 \leftarrow \{v_0\}$;
\STATE $U \leftarrow V \setminus \{v_0\}$;
\STATE $depth_T[.] \leftarrow $ an array of size $n$ that stores a depth for each vertex in a tree, filled with -1;
\STATE $bestNeighbor[.] \leftarrow $ an array that stores the best neighbor in $V_0$ for each vertex in $U$;
\STATE $wBestNeighbor[.] \leftarrow $ an array that stores the total power increase if the vertex will be connected with its best neighbor;
\STATE $depth(v_0) \leftarrow 0$;
\STATE $T \leftarrow (v_0, \emptyset)$;
\IF {$D$ is odd}
\STATE $v_1 \leftarrow ChooseSecondCenter(v_0)$;
\STATE $depth_T(v_1) \leftarrow 0$;
\STATE Add a vertex $v_1$ and an edge $(v_0, v_1)$ to $T$;
\ENDIF

\STATE $V_0 \leftarrow V_T$;

\FORALL {$u \in U$}
\STATE $bestNeighbor(u) \leftarrow \textsf{argmin}_{v \in V_T}\{Power(u, v)\}$;
\STATE $wBestNeighbor(u) \leftarrow Power(u, bestNeighbor(u))$;
\ENDFOR

\WHILE {$U$ is not empty}

\STATE $u \leftarrow ChooseEachVertex(U, V_0, wBestNeighbor)$;
\STATE Add a vertex $u$ and an edge $(u, bestNeighbor(u))$ to $T$;
\STATE $depth_T(u) \leftarrow depth_T(bestNeighbor(u) + 1)$;

\STATE $Power_T(u) \leftarrow Power(u, bestNeighbor(u))$;
\STATE $Power_T(bestNeighbor(u)) \leftarrow \max\{Power_T(bestNeighbor(u)), Power(u, bestNeighbor(u))\}$;

\STATE $U \leftarrow U \setminus \{u\};$

\FORALL {$v \in U$}

\STATE $w \leftarrow Power(bestNeighbor(u), v) + \max\{0, Power(bestNeighbor(u), v) - Power_T(v)\}$;

\IF {$w < wBestNeighbor(v)$}
\STATE $wBestNeighbor(v) \leftarrow w$;
\STATE $bestNeighbor(v) \leftarrow bestNeighbor(u)$;
\ENDIF

\ENDFOR

\IF {$depth_T(u) < \lfloor D / 2 \rfloor$}

\STATE $V_0 \leftarrow V_0 \cup \{u\}$;

\FORALL {$v \in U$}

\STATE $w \leftarrow Power(u, v) + \max\{0, Power(u, v) - Power_T(v)\}$;

\IF {$w < wBestNeighbor(v)$}
\STATE $wBestNeighbor(v) \leftarrow w$;
\STATE $bestNeighbor(v) \leftarrow u$;
\ENDIF

\ENDFOR

\ENDIF

\ENDWHILE

\IF {$W(T) < W^*$}
\STATE $W^* \leftarrow W(T)$;
\STATE $T^* \leftarrow T$;
\ENDIF
\ENDFOR
\STATE \textbf{return} $T^*$;
\end{algorithmic}
\caption{Prim-Like Heuristic} \label{alg:plh}
\end{algorithm}
\subsubsection{Min-Power Center-Based Tree Construction.}
The first algorithm based on PLH is Min-Power Center-Based Tree Construction (MPCBTC) which is similar to the Center-Based Tree Construction \cite{Julstrom09} for BDMST. In this algorithm, $ChooseFirstCenters$ chooses each vertex, that is, the algorithm starts $n$ times with each vertex of $V$ selected as a center. The method $ChooseSecondCenter(v_0)$ returns the vertex $v_1 = \textsf{argmin}_{v \in V \setminus \{v_0\}}{Power_T(v, v_0)}$. And, finally, the method $ChooseEachVertex(U, V_0, wBestNeighbor)$ finds such vertex $u \in U$ that $wBestNeighbor(u)$ is minimum. CBTC is known to perform worse with decrease of maximum hops number and increase of the points density, since the nodes that lie far from a center (let's call them \emph{far nodes}) often have the maximum allowable depth and, therefore, once added, they cannot be connected with any other node. And, for this reason, far nodes cannot be connected with any node in their proximity without violating the hops restriction, and they are forced to be connected with a tree by long arcs. Obviously, in MPCBTC, as well as in CBTC, the closest to the center nodes are added sooner, and in a case of large density and small $D$ MPCBTC will have the same disadvantage as CBTC: far nodes will be connected with a tree via long edges. Due to this fact solution obtained by MPCBTC should appear extremely inefficient for the cases when $n$ is large and $D$ is small. The computational complexity of MPCBTC is $O(n^3)$ since it is repeated $n$ times for each vertex chosen as the center, and each iteration requires $O(n^2)$ time.

\subsubsection{Min-Power Randomized Tree Construction.}
One simple approach aimed to overcome the mentioned disadvantage of CBTC is Randomized Tree Construction (RTC) proposed in \cite{Julstrom09}. As well as CBTC, RTC chooses a center vertex (or two center vertices if $D$ is odd), then it iteratively chooses a vertex outside a tree and connects it with some vertex in a tree. But in contrast to MPCBTC, each time the vertex is chosen at random. The process is repeated $n$ times, and the best tree is returned. We adapted this algorithm to MPBHSCP. Let's call the obtained heuristic as Min-Power Randomized Tree Construction (MPRTC). Since this algorithm is also based on PLH, the only parts that should be mentioned are the special implementations of the subroutines $ChooseFirstCenters$, $ChooseSecondCenter$, and $ChooseEachVertex$, which are extremely simple in this case: the method $ChooseFirstCenters$ $n$ times chooses a vertex $v \in V$ at random, as well as it is done in RTC \cite{Julstrom09}. The both methods $ChooseSecondCenter$ and $ChooseEachVertex$ choose a vertex $v \in U$ at random (where $U$ is a set of non-tree vertices, see Algorithm \ref{alg:plh}). This circumstance theoretically should cause better results of MPRTC comparing with MPCBTC on high-dense graphs constructed on uniformly spread set of points, because on each step of MPRTC the constructed partial solution consists of random subset of $V$. If $D$ is not too small, then the positions of the backbone vertices are also uniformly spread on each step, and therefore, on average, the weight of added edge should be rather small after some appropriate number of steps. Because of the fact that MPRTC is repeated $n$ times with different randomly chosen center, its total computational complexity is $O(n^3)$.

\subsubsection{Min-Power Center-based Least Sum-of-Costs.}
Another greedy algorithm for BDMST was proposed in \cite{Patvardhan2015}, it is called Center-based Least Sum-of-Costs. In similar manner to CBTC and RTC, it constructs a tree iteratively adding a vertex and an edge to the current tree. The difference of this algorithm from the mentioned above heuristics is that it chooses a vertex outside a tree with the minimum sum of costs of edges with other non-tree vertices. We employed a similar strategy and called the obtained algorithm Min-Power Center-based Least Sum-of-Costs (MPCBLSoC). But instead of minimizing the sum of the edge weights, we minimized the sum of the power costs in a star-like subgraph with a center in a given vertex what is more suitable for MPBHSCP. As well as the methods described above, MPCBLSoC is based on PLH. In this case, the methods $ChooseFirstCenters$, $ChooseSecondCenter$, and $ChooseEachVertex$ have the same implementation: given an already constructed partial tree $T$, there is selected a such vertex $v \in V \setminus V_T$, that a star graph on remaining vertices rooted in $v$ has minimum total power. The algorithm that chooses the \emph{best} star graph center is called $FindBestStarCenter$, and its pseudo-code is given in Algorithm \ref{alg:fc}. Thus, from the one hand, since $ChooseFirstCenters$ returns a single vertex, the algorithm MPCBLSoC contains the only iteration. But, from the other hand, $FindBestStarCenter$ runs in time $O(n^2)$, and, therefore, the total computational complexity of MPCBLSoC is $O(n^3)$.

\begin{algorithm}
\begin{algorithmic}
\STATE Input: $U \subset V$;
\STATE Output: $center \in U$;

\STATE $center \leftarrow \emptyset$;
\STATE $minCost \leftarrow \infty$;

\FORALL {$u \in U$}

\STATE $leavesCostSum \leftarrow 0$;
\STATE $centerCost \leftarrow 0$;

\FORALL {$v \in U \setminus u$}
\STATE $leavesCostSum \leftarrow leavesCostSum + Power(u, v)$;
\STATE $centerCost \leftarrow \max(Power(u, v), centerCost)$;
\ENDFOR

\IF {$centerCost + leavesCostSum < minCost$}
\STATE $center \leftarrow u$;
\STATE $minCost \leftarrow centerCost + leavesCostSum$;
\ENDIF
\ENDFOR

\STATE \textbf{return} $center$;

\end{algorithmic}
\caption{FindBestStarCenter} \label{alg:fc}
\end{algorithm}

\subsection{Min-Power Center-based Recursive Clustering}

Authors of \cite{Nghia2008} suggest another greedy heuristic called Center-based Recursive Clustering (CBRC) for BDMST. This algorithm
 starts with a spanning star tree rooted in the center, chosen in such a way that the sum of edge weights is minimum. Then the leaves, whose depth is less than $\lfloor D/2 \rfloor$, are iteratively reorganized into a cluster with a center in some node. On each iteration, the leaves are reattached to a center if this improves solution and the restriction on the number of hops is held.  We called our implementation for MPBHSCP Min-Power Center-based Recursive Clustering (MPCBRC). As a center choosing subroutine the previously described algorithm  $FindBestStarCenter$ is used. The pseudo-code of MPCBRC is presented in Algorithm \ref{alg:cbrc}. Each iteration of the algorithm takes $O(n^2)$ operations because of the complexity of $FindBestStarCenter$, and, since there are $O(n)$ iterations, the algorithm runs in time $O(n^3)$.

\begin{algorithm}
\begin{algorithmic}
\STATE $v_0 \leftarrow FindBestStarCenter(V)$;
\STATE $V_0 \leftarrow \{v_0\}$;
\STATE $T \leftarrow $ a star graph rooted in $v_0$;
\STATE $U \leftarrow V \setminus \{v_0\}$;
\STATE $depth_T[.] \leftarrow $ an array of size $n$ that stores a depth for each vertex in a tree, filled with -1;
\STATE $depth(v_0) \leftarrow 0$;
\IF {$D$ is odd}
\STATE $v_1 \leftarrow FindBestStarCenter(V_0)$;
\STATE $depth_T(v_1) \leftarrow 0$;
\STATE Add a vertex $v_1$ and an edge $(v_0, v_1)$ to $T$;
\ENDIF

\WHILE {$U$ is not empty}

\STATE {$U_0 \leftarrow \{v \in U: depth_T(v) < \lfloor D/2 \rfloor \}$}
\STATE $center \leftarrow FindBestStarCenter(U_0)$;

\IF {$center == \emptyset$}
\STATE \textbf{break};
\ENDIF

\STATE $U \leftarrow U \setminus \{center\}$;

\FORALL {$u \in U$}
\STATE Set $powerIncrease \leftarrow \{$power increase after reassigning a parent of $u$ from $Parent_T(u)$ to $center\}$;
\IF {$powerIncrease < 0$}
\STATE $T \leftarrow (T \setminus \{(u, Parent_T(u))\}) \cup \{(u, center)\}$;
\STATE $depth_T(u) = depth_T(center) + 1$;
\ENDIF
\ENDFOR

\ENDWHILE

\end{algorithmic}
\caption{Min-Power Center-based Recursive Clustering} \label{alg:cbrc}
\end{algorithm}

\subsection{Min-Power Quadrant Center-based Heuristic}

One of the most efficient heuristics applied to BDMST in planar Euclidian case with uniformly distributed vertices consists in recursive splitting the given region into equal parts (\emph{quadrants}) and search of their centers \cite{Patvardhan2015}. We implemented a variant of the similar approach for MPBHMSCP and called it Min-Power Quadrant Center-based Heuristic (MPQCH). The pseudo-code of this algorithm is given in Algorithm \ref{alg:qbh}. As well as in some of the previous heuristics, it starts with choosing a center by the algorithm $FindBestStarCenter$. But this time in order to reach central symmetry we choose the only start center despite the parity of $D$. Then inside the main loop the region is iteratively split into the squared cells of equal size. For each cell, its center is chosen by the algorithm $FindBestStarCenter$ and then it is added to the tree with an edge that connects it with a center of a previous iteration's cell that contains it, or with $v_0$ at the first iteration. At each iteration the number of cells four times greater than the number of cells in the previous iteration, that is, each cell consists of four cells of the next iteration. At each iteration the height of a constructed tree is increased by 1, and, since $stepsCount$ is bounded by $\lfloor D/2 \rfloor$, the diameter constraint is not violated.

In our implementation, for the speed purposes, a regular rectangular grid of size $qsize \times qsize$ is initially set on the given region, and a corresponding grid cell is assigned to each vertex.  Then, due to this grid, during the main loop the subset of vertices that belong to each cell $c \in C$ are found in constant time. Actually, $qsize$ is a parameter of the algorithm, and the greater value of $qsize$ allows to obtain better solution but increases the running time. The computational complexity of the algorithm is $O(qsize^2 + \min\{\lfloor D/2 \rfloor, \log(qsize)\} n^2)$.

\begin{algorithm}
\begin{algorithmic}
\STATE $v_0 \leftarrow FindBestStarCenter(V)$;
\STATE $T \leftarrow (\{v_0\}, \emptyset)$;
\STATE $U \leftarrow V \setminus \{v_0\}$;
\STATE Construct rectangular grid of size $qsize \times qsize$ on a given square;
\STATE $stepsCount \leftarrow \min(\lfloor D/2 \rfloor, \log_2(qsize))$;
\STATE $cellCenter$ --- an array of size $n$ that stores a cell center for each vertex;
\STATE Fill $cellCenter$ with $v_0$ (initially the whole square is a single cell and the root is a center);

\FORALL {$step \in \{1, ..., stepsCount\}$}
\STATE Split grid into $2^{step} \times 2^{step}$ cells $C$ of equal size;

\FORALL {$c \in C$}
\STATE $U_c \leftarrow $ vertices of $U$ located in $c$;
\STATE $center \leftarrow FindBestStarCenter(U_c)$;
\STATE $T \leftarrow T \cup \{(bestCenter, cellCenter(center))\}$;
\STATE $U \leftarrow U \setminus \{center\}$;
\FORALL {$u \in U_c \setminus \{center\}$}
\STATE $cellCenter(u) \leftarrow center$;
\ENDFOR
\ENDFOR
\ENDFOR

\end{algorithmic}
\caption{Min-Power Quadrant Center-based Heuristic} \label{alg:qbh}
\end{algorithm}

\begin{algorithm}[!hb]
\begin{algorithmic}
\STATE $v_0 \leftarrow FindBestStarCenter(V)$;
\STATE Construct spanning tree $T$ rooted in $v_0$ by IPP;
\STATE $V_0 \leftarrow \{v_0\}$;
\STATE $U \leftarrow V \setminus \{v_0\}$;
\STATE $depth(v_0) \leftarrow 0$;
\IF {$D$ is odd}
\STATE $v_1 \leftarrow $ most remote neighbor of $v_0$ in $T$;
\STATE $depth_T(v_1) \leftarrow 0$;
\STATE Add a vertex $v_1$ and an edge $(v_0, v_1)$ to $T$;
\ENDIF
\STATE Calculate the values of $depth_T$;

\STATE $U \leftarrow \{v \in V \setminus \{s\}: depth_T(v) > h\}$;

\WHILE {$U$ is not empty}

\STATE $bestChild \leftarrow \emptyset$;
\STATE $bestParent \leftarrow \emptyset$;
\STATE $minPowerIncrease \leftarrow \infty$;
\STATE Mark all vertices in $U$ as not considered;

\FORALL {$u \in U$}

\STATE $C \leftarrow \{u\} \cup \{v \in V: depth_T(v) > 1 \;\&\; v \;$ is predecessor of $ u $ in $ T\}$

\FORALL {$c \in \{$not considered elements of $C$\}}
\IF {$c$ is considered}
\STATE \textbf{continue};
\ENDIF
\STATE Mark $c$ as considered;
\STATE $P \leftarrow \{v \in V: depth_T(v) < \min(\lfloor D/2 \rfloor-1, depth_T(c) - 1)\}$;
\FORALL {$p \in P$}
\STATE $powerIncrease \leftarrow $maximum power costs change of vertices $c$, $Parent_T(c)$, and $p$ after assigning $p$ as a parent of $c$ in $T$;

\IF {$powerIncrease < minPowerIncrease$}
\STATE $minPowerIncrease \leftarrow powerIncrease$;
\STATE $bestChild \leftarrow c$;
\STATE $bestParent \leftarrow p$;
\ENDIF

\ENDFOR
\ENDFOR
\ENDFOR

\STATE $T \leftarrow T \setminus (\{(bestChild, Parent_T(bestChild))\}) \cup \{(bestChild, bestParent)\}$;

\STATE Decrease $Level_T$ for all the vertices in the branch rooted in $bestChild$ by $Level_T(bestChild) - depth_T(bestParent) - 1$;

\STATE $U \leftarrow U \setminus \{v \in U: depth_T(v) \leq \lfloor D/2 \rfloor\}$;
\ENDWHILE

\end{algorithmic}
\caption{Min-Power Iterative Refinement} \label{alg:ir}
\end{algorithm}

\subsection{Min-Power Iterative Refinement}

Another good approach for building spanning tree with bounded diameter is, first, construction a tree without restriction on diameter and, after that, iteratively decrease depths of vertices until the restriction on diameter is satisfied. The iterative algorithm that reduces the diameter of an input spanning tree for BDMST has been proposed in \cite{Deo2000}. We propose the heuristic for MPBHSCP called Min-Power Iterative Refinement (MPIR), which is based on the similar idea. The pseudo-code of this algorithm is presented in Algorithm \ref{alg:ir}. At first, a center $v_0$ is chosen by the $FindBestStarCenter$ subroutine. Then, a near-optimal solution for an unbounded problem  rooted in $v_0$ is constructed by IPP \cite{CNSCL03}. If $D$ is odd, then the most remote neighbor of $v_0$ in $T$ is selected as second center.  The algorithm works with a set of vertices $U$ whose depth exceeds $\lfloor D/2 \rfloor$. For each $u \in U$ the best removing of an edge from the path from $u$ to $v_0$ and subsequent
adding another edge that decreases a depth of $u$ and increases the $W(T)$ at least is found. The best of such edge exchanges among all vertices of $U$ is performed. After each modification of a tree depth of some vertices in $U$ may be decreased, therefore, the vertices whose depth is less than $\lfloor D/2 \rfloor$ are removed then from $U$. The computational complexity of the algorithm is $O(n^3)$.

\section{Simulation}\label{sS}
We have implemented all the described algorithms in C++ programming language and run them on the data sets that are given in Beasley's OR-Library for Euclidian Steiner Problem (http://people.brunel.ac.uk/~mastjjb/jeb/orlib). These test cases present the random uniformly distributed points in the unit square. For the same dimension 15 different instances are provided. We tested 4 variants of dimension: $n = $ 100, 250, 500, and 1000, 15 instances for each. We also took different values of $D$ for each dimension. The experiment was launched on the Intel Core i5-4460 3.2GHz processor with 8Gb RAM.

\begin{table}[!hbtp]
\resizebox{0.75\textwidth}{!}
 {
\setlength{\tabcolsep}{1.5pt}
 \begin{minipage}{\textwidth}
 \setlength{\tabcolsep}{0.3em}
  \begin{tabular}{c|c|ccc|ccc|ccc|ccc|ccc|ccc}

  \multirow{2}{*}{n} & \multirow{2}{*}{D}
  &\multicolumn{3}{c|}{MPCBTC}
  &\multicolumn{3}{c|}{MPRTC}
  &\multicolumn{3}{c|}{MPCBLSoC}
  &\multicolumn{3}{c|}{MPCBRC}
  &\multicolumn{3}{c|}{MPQBH}
  &\multicolumn{3}{c}{MPIR}\\
& & av & err & time & av & err & time & av & err & time & av & err & time & av & err & time & av & err & time \\
\hline
100	&5	&8.17	&0.47	&0	&\textbf{3.6}	&0.13	&0	&8.8	&0.66	&0	&8.41	&0.79	&0	&5.04	&0.17	&0	&12.1	&0.48	&0\\
	&10	&3.38	&0.21	&0	&1.88	&0.06	&0	&3.5	&0.31	&0	&3.07	&0.29	&0	&2.06	&0.07	&0	&\textbf{1.84}	&0.14	&0\\
	&15	&1.87	&0.16	&0	&1.75	&0.07	&0	&1.62	&0.18	&0	&2.39	&0.17	&0	&2.06	&0.07	&0	&\textbf{1.19}	&0.05	&0\\
	&25	&0.92	&0.05	&0	&1.74	&0.07	&0	&0.96	&0.03	&0	&2.36	&0.17	&0	&2.06	&0.07	&0	&\textbf{0.89}	&0.02	&0\\
250	&10	&13.2	&0.93	&0.07	&\textbf{2.32}	&0.07	&0.09	&14.1	&1.56	&0.02	&6.98	&1.28	&0.02	&2.44	&0.04	&0	&5.41 & 0.43	&0.04\\
	&15	&8.17	&0.65	&0.08	&\textbf{2}	&0.03	&0.1	&7.94	&0.97	&0.02	&3.85	&0.29	&0.02	&2.47	&0.04	&0	&2.6	&0.46	&0.05\\
	&20	&4.3	&0.5	&0.08	&2.03	&0.05	&0.11	&3.49	&0.42	&0.02	&3.31	&0.25	&0.02	&2.47	&0.04	&0	&\textbf{1.48}	& 0.22	&0.04\\
	&40	&0.96	&0.05	&0.1	&2.03	&0.05	&0.11	&\textbf{0.91}	&0.02	&0.02	&3.32	&0.22	&0.02	&2.47	&0.04	&0	&1.04	& 0.26	&0.02\\
500	&15	&26	&1.89	&0.65	&\textbf{2.26}	&0.03	&1	&26.6	&2.5	&0.14	&6.24	&0.78	&0.14	&2.62	&0.03	&\textbf{0.03}	&5.1	&0.49	&0.33\\
	&30	&6.37	&0.52	&0.78	&2.2	&0.04	&1.04	&4.23	&0.61	&0.14	&3.88	&0.27	&0.15	&2.62	&0.03	&\textbf{0.03}	&\textbf{1.41}	&0.17	&0.34\\
	&45	&1.87	&0.19	&0.8	&2.2	&0.04	&0.94	&1.1	&0.09	&0.13	&3.89	&0.25	&0.15	&2.62	&0.03	&\textbf{0.03}	&\textbf{1.04}	&0.11	&0.23\\
	&60	&0.91	&0.04	&0.95	&2.2	&0.04	&1.04	&0.89	&0.01	&0.14	&3.88	&0.27	&0.16	&2.62	&0.03	&\textbf{0.04}	& \textbf{0.857}	&0.04	&0.16\\
1000	&20	&50.4	&1.98	&6.27	&\textbf{2.45}	&0.04	&13.5	&49.4	&3.01	&1.13	&6	&0.57	&1.23	&2.81	&0.02	&\textbf{0.15}	&5.26	&0.43	&2.94\\
	&40	&14.6	&1.39	&8.4	&2.43	&0.03	&14.6	&8.87	&1.02	&1.16	&4.52	&0.32	&1.32	&2.81	&0.02	&\textbf{0.16}	&\textbf{1.52}	&0.16	&3.07\\
	&60	&4.02	&0.33	&9.88	&2.42	&0.03	&15.2	&1.25	&0.09	&1.16	&4.52	&0.32	&1.34	&2.81	&0.02	&\textbf{0.16}	&\textbf{1.12}	&0.11	&2.43\\
	&100	&\textbf{0.81}	&0.02	&11.7	&2.44	&0.02	&14.1	&0.9	&0	&1.15	&4.52	&0.32	&1.29	&2.81	&0.02	&\textbf{0.16}	&\textbf{0.85}	&0.05	&1.25\\

  \end{tabular}
\medskip
\end{minipage}}
  \caption{Comparison of the experiment's results obtained by different heuristics.}
 \label{table:expRes}
\end{table}

For the algorithm MPQCH we chose $qsize = n$ since such value does not slow down the algorithm match, while the solution quality is significantly greater than in the case $qsize = \sqrt{n}$.

The results of the experiment are presented in Table \ref{table:expRes}. For each algorithm and each tested combination of $n$ and $D$ the average objective value (av), average time in seconds (t), and standard deviation (err) are shown. In average, when the diameter bound is low, the best solution is constructed by MPRTC. With large values of $D$ MPIR constructs the best solution. Note that MPÑBTC and MPCBLCoS results are very poor when $D$ is small, but with large values of $D$ their average objective values are close to minimum. MPÑBTC and MPRTC appeared to be the most time consuming on large dimension cases, while MPQCH always runs significantly faster than other algorithms. Besides, MPQCH performance almost  does not depends on $D$. Most probably this is because the maximum diameter of the constructed solution is much less than $D$, --- this gives us a possibility for the further improvements of this algorithm.

\begin{figure}[!hbtp]
\centering
\subfloat[\label{fig:expres_CBTC}MPCBTC. $W(T) = 7.37$]{\includegraphics[width=0.33\textwidth]{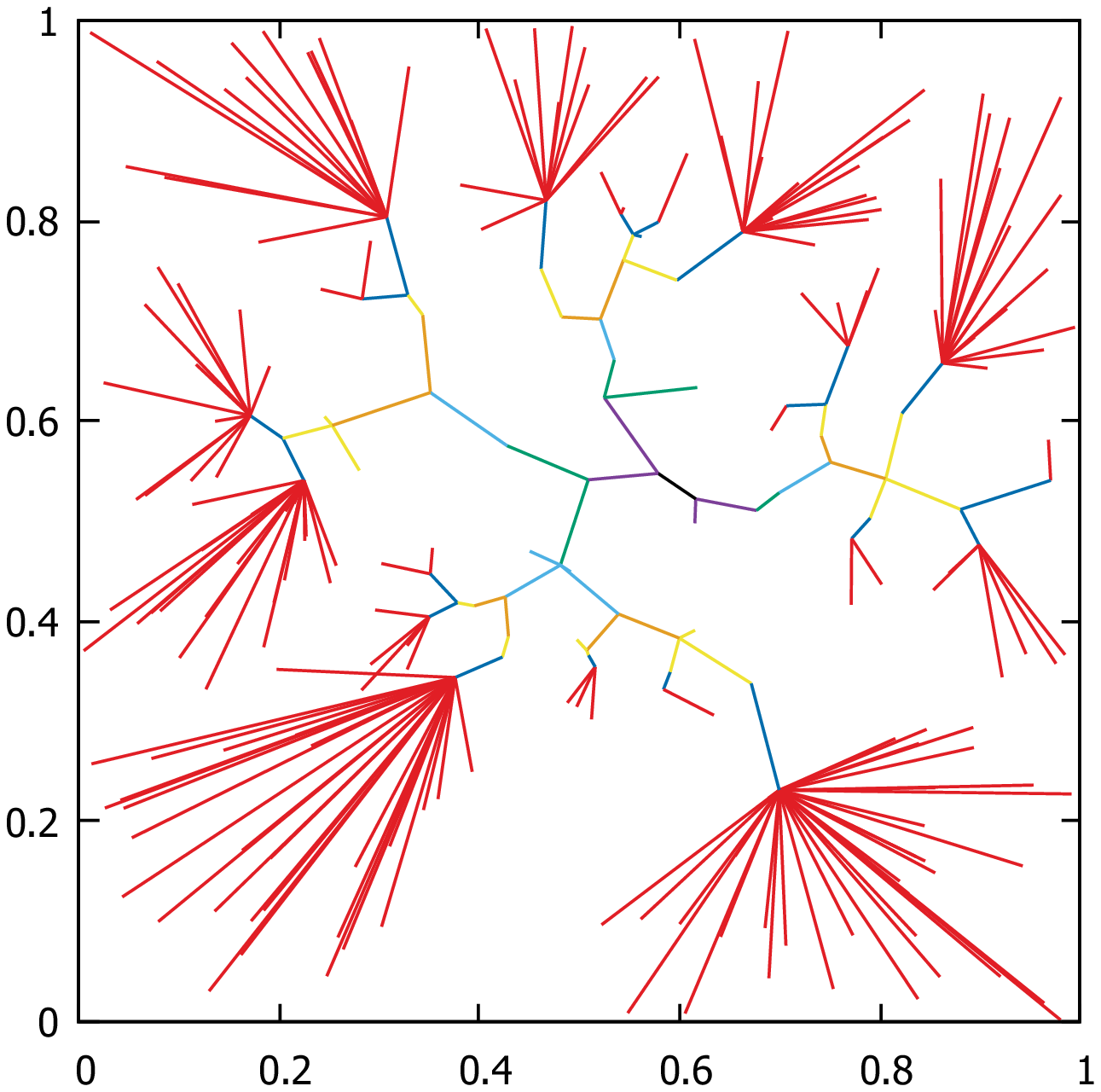}} \hfill
\subfloat[\label{fig:expres_RTC}MPRTC. $W(T) = 1.99$]{\includegraphics[width=0.33\textwidth]{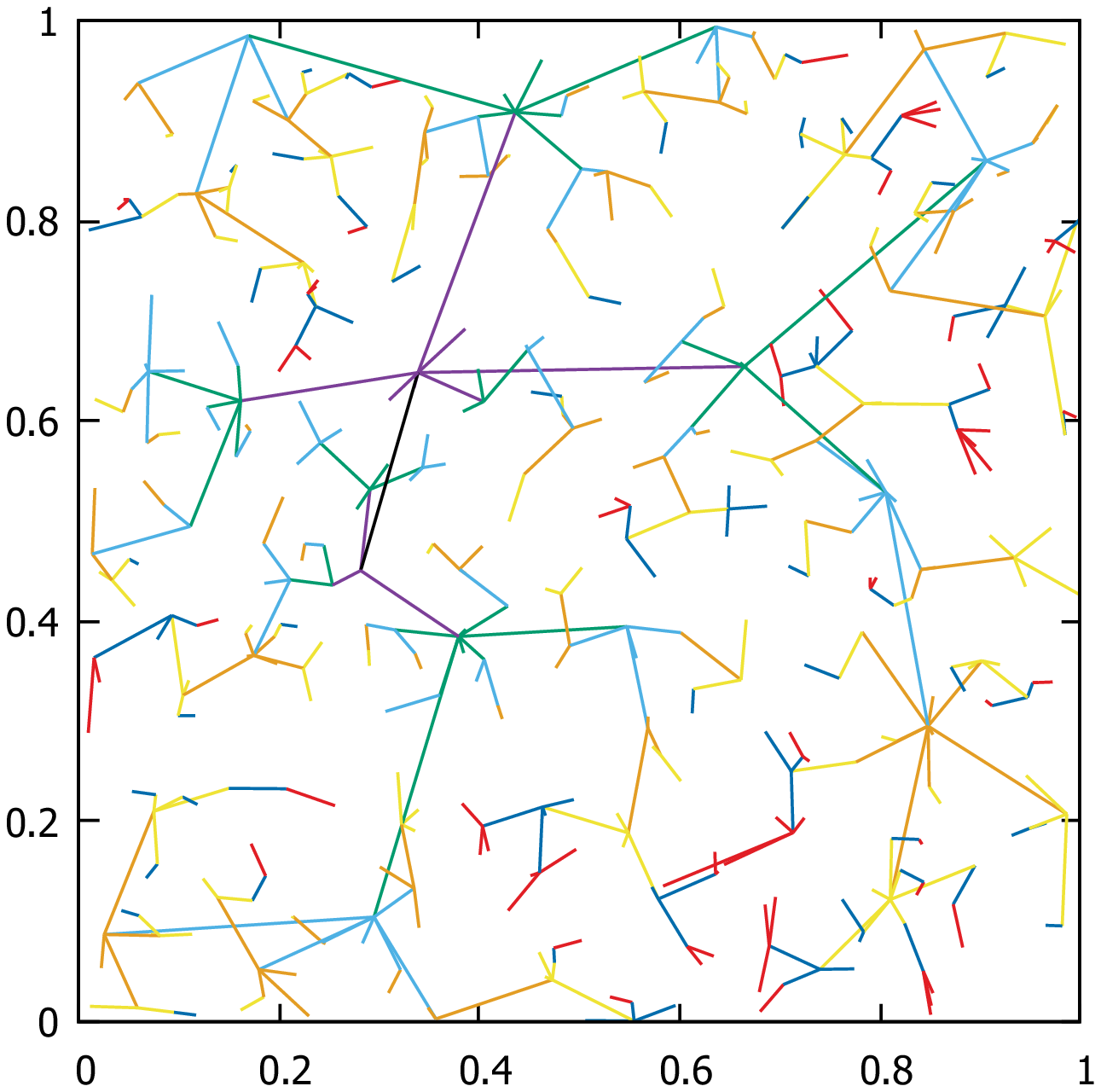}} \hfill
\subfloat[\label{fig:expres_CBLSoC}MPCBLSoC. $W(T) = 4.78$]{\includegraphics[width=0.33\textwidth]{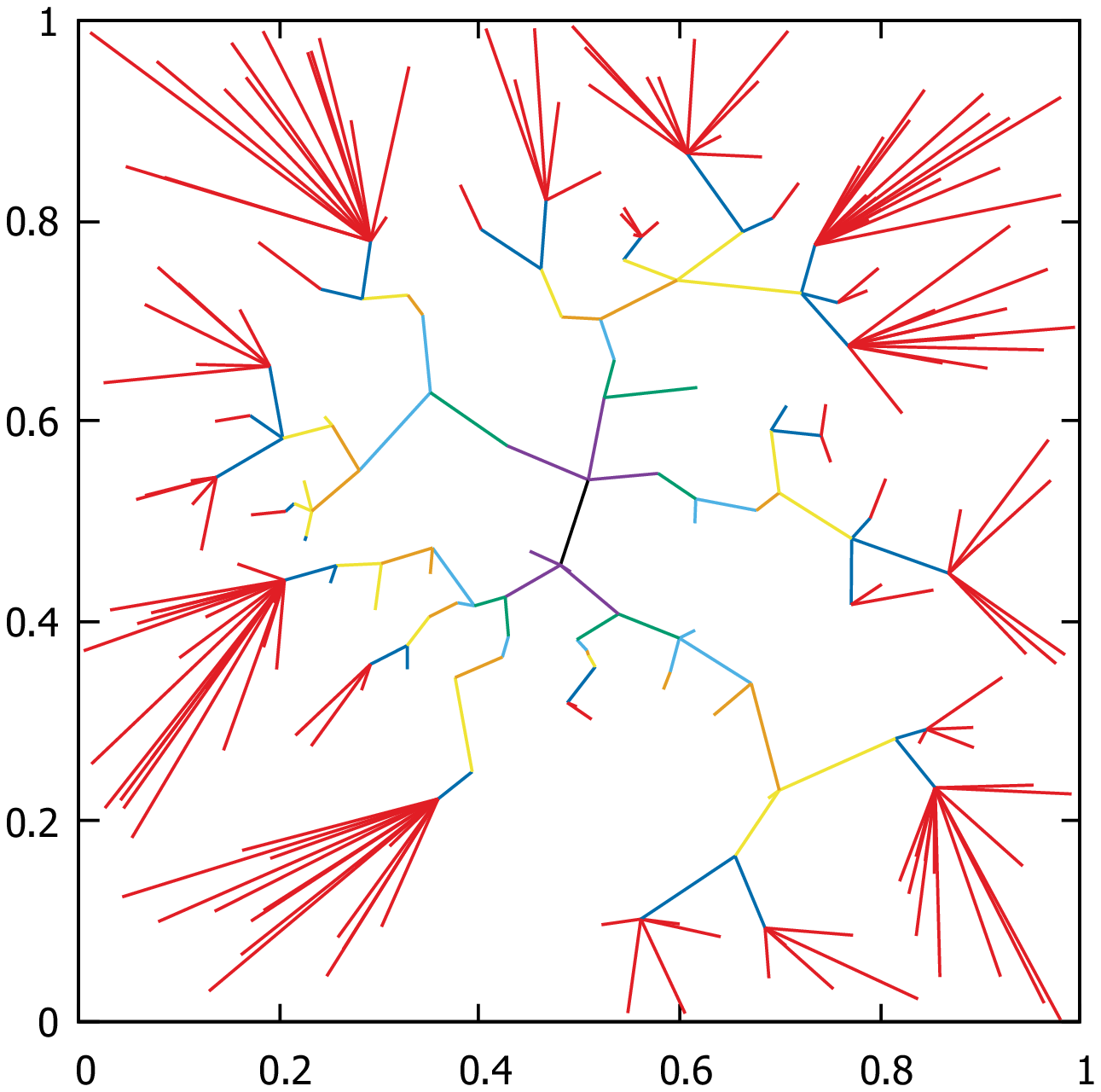}} \hfill
\subfloat[\label{fig:expres_CBRC}MPCBRC. $W(T) = 3.59$]{\includegraphics[width=0.33\textwidth]{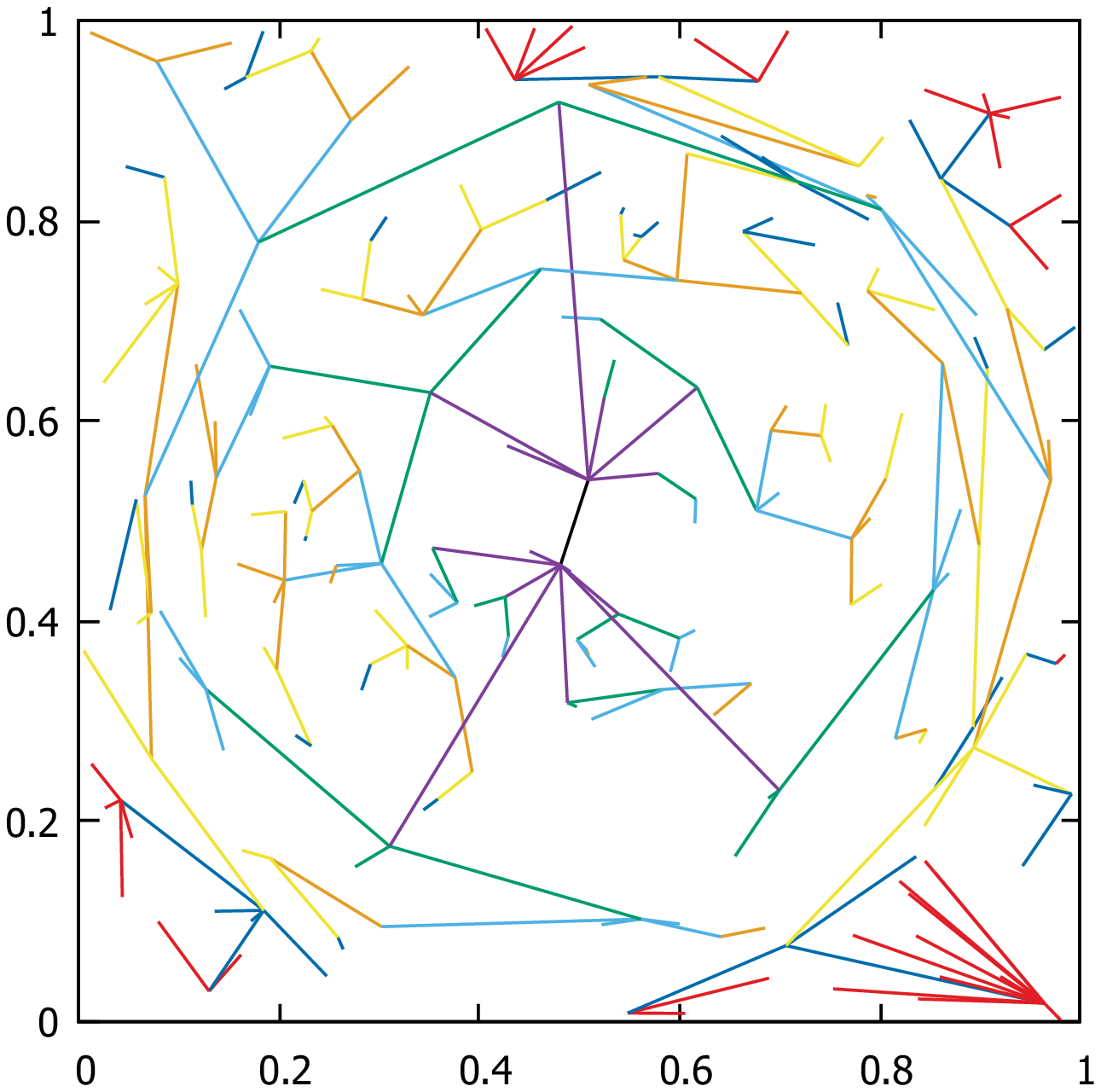}} \hfill
\subfloat[\label{fig:expres_IR}MPIR. $W(T) = 2.44$]{\includegraphics[width=0.33\textwidth]{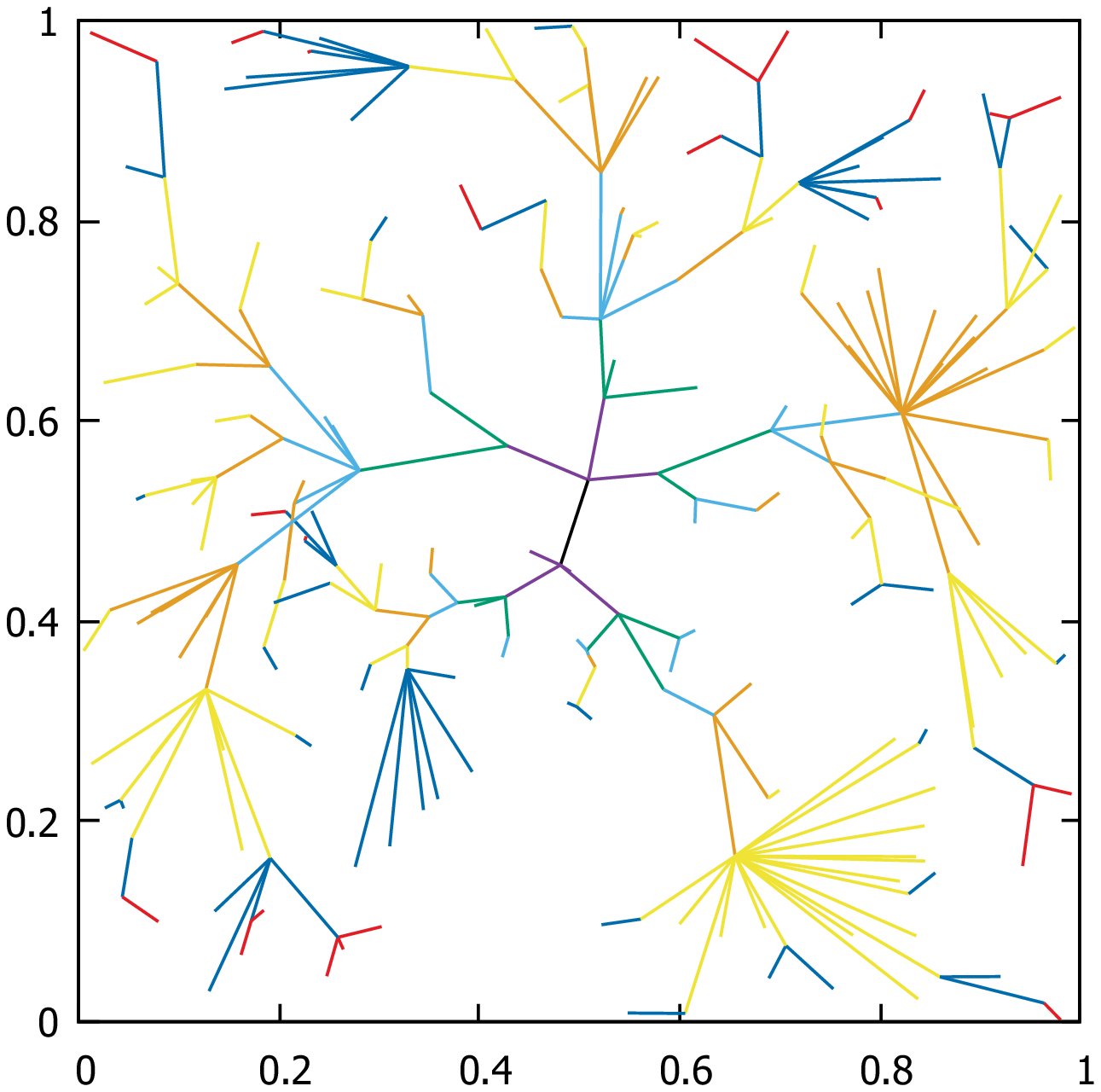}} \hfill
\subfloat[\label{fig:expres_QCH}MPQCH. $W(T) = 2.35$]{\includegraphics[width=0.33\textwidth]{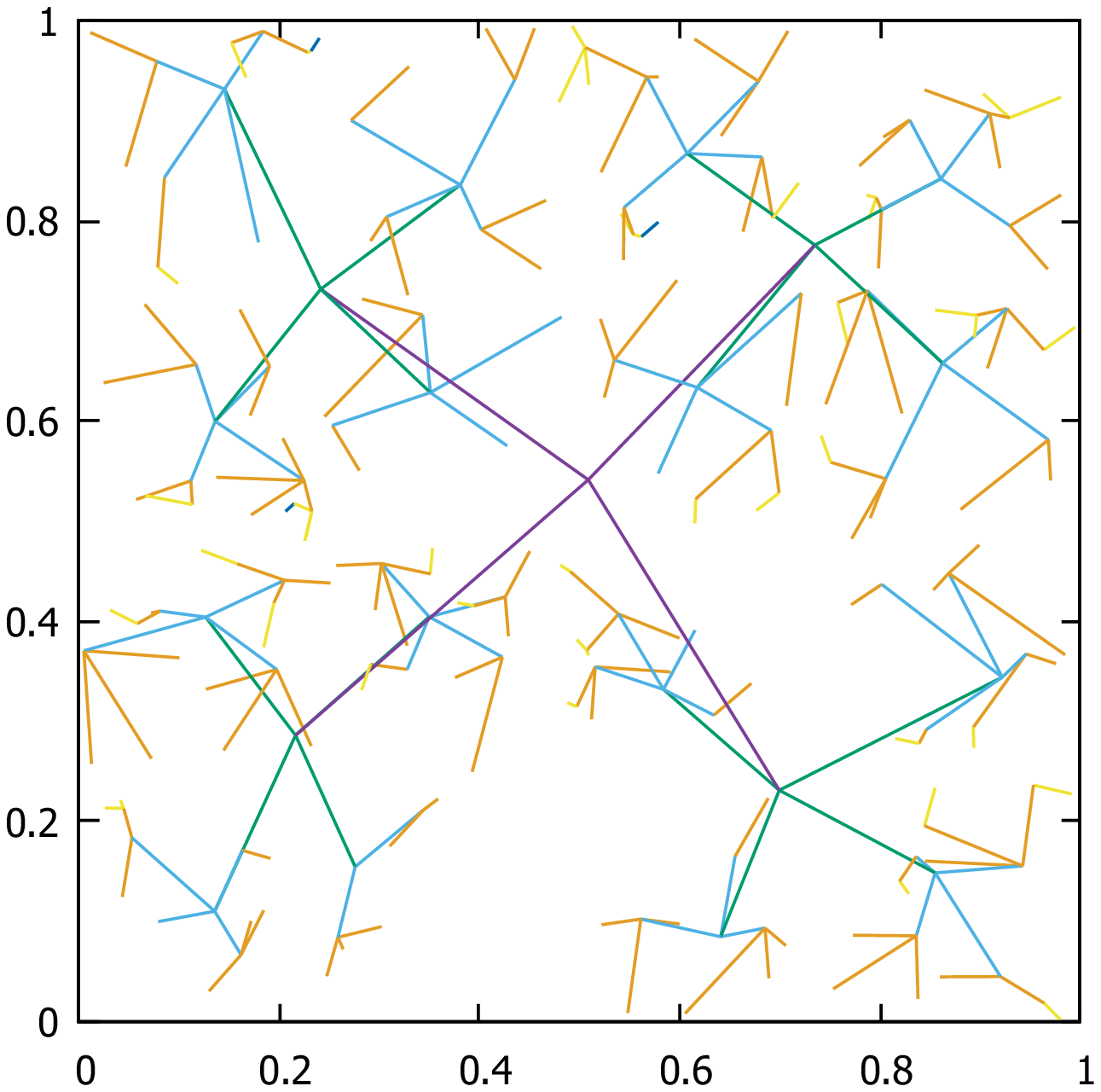}} \hfill
\caption{Algorithms results on the same instance. $D = 15, n = 250$} \label{fig:expres}
\end{figure}

As an illustration, we also present in Fig. \ref{fig:expres} the solutions that were obtained by different algorithms on the same instance when $D = 15, n = 250$. For the convenience, the edges that remote from a center by an equal distance (i.e., hops count) are colored in the same color. Since the diameter bound is odd in this case, there are two centers (connected by a black edge) in solutions constructed by all algorithms except MPQCH, which always builds a tree with the only center. The difference in the behaviour of the algorithms is seen in these pictures. The diameter bound is still not enough for MPCBTC and MPCBLSoC to construct good solutions: in both cases the backbone is too small and there are many leaves far from a center that are coincident with long edges (colored in red). MPCBRC constructs a tree with a lot of long edges in backbone, since the backbone vertices are always chosen as center of the current set of leaves during the tree construction. MPIR result contains a lot of vertices with large degree that are coincident with rather long edges, that slightly deteriorate solution. The remained two algorithms, MPRTC and MPQCH, that performed the best, have the following common features: (1) the number of vertices increases with increasing of their depths; (2) the average edge weight decreases with increase of the depth. MPRTC always chooses a vertex at random, and, in average, the distance to the closest in-tree vertex becomes less while the constructed tree size grows. MPQCH constructs a tree whose backbone vertices are located close to the quadrants geometric centers. Note that MPQCH built a tree with maximum depth equal 6 while the depth upper bound is 7. This allows to improve solution in this case: each of the longest edges that connect a center with its four children could be replaced by two shorter edges with intermediate vertex that is located close to edge's geometric center. We assume that such modification will significantly improve the solution, and we plan to implement it in future.

\section{Conclusion}\label{sC}
In this paper, the NP-hard Min-Power Bounded Hops Symmetric Connectivity Problem was considered. We proposed six different constructive heuristics for its approximation solution. As main ideas of our approaches, we used some of the known heuristics that were previously developed for BDMST. We implemented all the proposed algorithms and conducted the numerical experiment on different randomly generated test instances. The simulation shows that in cases with large diameter the algorithm MPIR yields the best results, while the usage of MPRTC is more preferable when the diameter is small. If one needs to obtain a solution of rather good quality in shortest time, then MPQCH could be the best choice. Besides, the experiment results show that MPQCH can be significantly improved. In future we plan to develop different variants of local search and other metaheuristics that appeared to be efficient for BDMST, such as variable neighborhood search, genetic algorithm, and ant colony optimization, where the trees obtained by different algorithms proposed in this paper will serve as start solutions.

\end{document}